\documentstyle[aps,epsf,prc]{revtex}

\begin{document}
\title{Polarization phenomena for heavy meson production in NN-collisions near threshold}
\author{Michail P. Rekalo and  Egle Tomasi-Gustafsson}
\address{\it DAPNIA/SPhN, CEA/Saclay, 91191 Gif-sur-Yvette Cedex, 
France}
\maketitle
\begin{abstract}

Polarization phenomena are essential for the understanding of elementary processes, and are typically large in the threshold region. The selection rules with respect to P-parity, angular momentum, isotopic spin and the Pauli principle allow to parametrize in a model independent way, the 
spin structure of the matrix elements for the near-threshold heavy meson production in NN-collisions 
in terms of a small number of partial amplitudes. Such parametrization is very powerful for the analysis of polarization phenomena and  possible  $t-$exchange mechanisms.
\end{abstract}
\section{Introduction}
The study of elementary processes such as heavy meson production in NN-collisions is very important for building reliable microscopic models and for the understanding of nucleon-nucleus and nucleus-nucleus collisions. Polarization phenomena give the possibility to test symmetry properties and to disentangle the problems related to the hadron structure or to the reaction mechanism. We mention, as examples, the possibility to determine experimentally the P-parity of the K-meson, and the $\Lambda N$ scattering lengths, through the reaction $p+p\to p+\Lambda +K$, at threshold. 

The near threshold region, which is characterized by the production of final particles in relative S-state, is particularly interesting, as:
\begin{itemize}
\item it is, in general, quite wide for $V^0$-meson production, and can be identified by the behavior of experimental observables such as angular distributions and polarization observables,
\item polarization phenomena are very peculiar, either they vanish or they are large (in absolute value), reaching their limiting values, especially in $pp$-collisions
\item  isotopic effects are large: for $np$-collisions the cross section is larger than for $pp$-collisions and polarization phenomena are more rich.
\end{itemize}

In this contribution we will illustrate very briefly a formalism \cite{ref1}, based on fundamental properties of the strong interaction, and give predictions, in model independent way, on reactions and observables which can be accessed at COSY.

\section{Spin structure of threshold matrix elements and polarization phenomena}

The formalism developed here, applies to vector \cite{ref2}, scalar, pseudoscalar meson \cite{ref3} as well as to strange  particle production \cite{ref4}. It is based on the symmetry properties of the strong interaction, such as the Pauli principle, the isotopic invariance, the P-invariance, and it is  particularly simplified in the threshold region, where all final particles are produced in $S$-state. 

\subsection{Vector meson production}
In the general case the spin structure of the matrix element for the process 
$N+N\to N+N+V$ is described by a set of 48 independent complex amplitudes, which are functions of five kinematical variables. The same reaction, in coplanar kinematics, is described by 24 amplitudes, functions of four variables. In collinear kinematics the number of independent amplitudes is reduced to 7 and the description of this reaction is further simplified in case of threshold $V$-meson production, where one amplitude describes the $pp$-reaction and two amplitudes the $np$-reaction.
the threshold process $p+p \rightarrow p+p+V^0$ is characterized by a single partial transition: $
S_i=1,~\ell_i=1~\to ~{\cal J}^{ P}=1^- \to S_f=0$, 
where $S_i$ ($S_{f}$) is the total spin of the two protons in the initial (final) 
states and $\ell_i$ is the orbital momentum of the colliding protons.
In the CMS of the considered reaction the corresponding matrix element can be written as:
\begin{equation}
{\cal M}(pp)=2f_{10}(\tilde{\chi}_2~\sigma_y ~\vec{\sigma}
\cdot\vec  U^* \times\hat{\vec k }\chi_1)~(\chi^{\dagger}_4 \sigma_y\ \tilde{\chi}^{\dagger}_3 )
\label{eq:mpp},
\end{equation}
where $\chi_1$ and $\chi_2$ ( $\chi_3$ and $\chi_4$) are the
two-component spinors of the initial (final) protons;  $\vec  U$ is the three-vector of the $V$-meson polarization, $\hat{\vec k}$ is
the unit vector along the 3-momentum of the initial proton; $f_{10}$ is the S-wave partial amplitude, describing the triplet-singlet transition of the two-proton system in V-meson production.

In case of  $np$-collisions, applying the conservation of isotopic invariance for the strong interaction, two threshold partial transitions are allowed: $S_i=1,~\ell_i=1~\to ~{\cal J}^{ P}=1^- \to S_f=0,$ and $ S_i=0,~\ell_i=1~\to ~{\cal J}^{ P}=1^- \to S_f=1,$ with the following spin structure of the matrix element:
\begin{equation}
{\cal M}(np)=f_{10}(\tilde{\chi}_2~\sigma_y ~\vec{\sigma}
\cdot\vec  U^* \times\hat{\vec k }\chi_1)~(\chi^{\dagger}_4 \sigma_y  \tilde{\chi}^{\dagger}_3 )+f_{01}(\tilde{\chi}_2~\sigma_y\chi_1)(\chi^{\dagger}_4
\vec{\sigma}
\cdot\vec  U^* \times\hat{\vec k }\sigma_y \tilde{\chi}^{\dagger}_3 ),
\label{eq:mnp}
\end{equation}
where $f_{01}$ is the S-wave partial amplitude describing the singlet-triplet transition of the two-nucleon system in V-meson production. In the general case the amplitudes $f_{10}$ and $f_{01}$ are complex functions, depending on the energies $E$, $E'$ and $E_V$, where $E,(E')$ and $E_V$ are the energies of the initial (final) proton and of the produced $V-$meson, respectively. Note that $f_{10}$ is the common amplitude for $pp$- and $np$-collisions, due to the isotopic invariance of the strong interaction. 

All dynamical information is contained in the partial amplitudes $f_{01}$ and $f_{10}$, which are different for the different vector particles. On the other hand, some polarization phenomena have common characteristics, essentially independent from the type of vector meson. For example, vector mesons produced in $pp$- and $np$-threshold collisions are transversally polarized,
and the elements of the density matrix $\rho$ are independent from the relative values of the amplitudes $f_{01}$ and $f_{10}$: $\rho_{xx}=\rho_{yy}=\frac
{1}{2}$, $\rho_{zz}=0$. Therefore, the angular distribution has a  $\sin^2\theta_P$-dependence for the subsequent decay $V^0\to P+P$ (where $P$ is a pseudoscalar meson) and a $(1+\cos^2\theta)$-dependence  for the decay
$V^0\to \mu^+ +\mu^-$, where $\theta$ ($\theta_P$) is the angle between $\hat{\vec k}$ and the $\mu^-$ $(P)$ momentum (in the rest system of $V^0$). Possible deviations from this behavior have to be considered as an indication of the presence of higher partial waves in the final state. All other one-spin polarization observables, related to the polarizations of the initial or final nucleons, identically vanish, for any process of $V-$meson production.

The dependence of the differential cross section for threshold collisions of polarized nucleons (where the polarization of the final particles is not detected) can be parametrized as follows:

\begin{equation}
\displaystyle\frac{d\sigma}{d\omega}(\vec P_1,\vec P_2)=\left ( \displaystyle\frac{d\sigma}{d\omega}\right)_0
\left (1+{\cal A}_1 \vec P_1 \cdot \vec P_2 +{\cal A}_2 \hat{\vec k} \cdot\vec P_1\hat{\vec k} \cdot\vec P_2 \right ),
\label{eq:sig}
\end{equation}
where $\vec P_1 $ and $\vec P_2$ are the axial vectors of the beam and target nucleon polarizations, $d\omega$ is the element of phase-space for the three-particle final state. The spin correlation  coefficients  ${\cal A}_1$ and ${\cal A}_2$ are real and they are different for $pp$- and $np$- collisions:
\begin{itemize}
\item $\vec p+\vec p\to p+p+V^0$:  ${\cal A}_1(pp)=0$, ${\cal A}_2(pp)=1$.
\item $\vec n+\vec p\to n+p+V^0$:  
${\cal A}_1(np)=-\displaystyle\frac{|f_{01}|^2}{|f_{01}|^2+|f_{10}|^2},~~{\cal
A}_2(np)=\displaystyle\frac{|f_{10}|^2}{|f_{01}|^2+|f_{10}|^2}$,
\end{itemize}
with the following relations $-{\cal A}_1(np)+{\cal A}_2(np)=1$ and 
$0\le {\cal A}_2(np)\le 1$.

The ratio ${\cal R}$ of the total (unpolarized) cross section for $np$- and $pp$- collisions, taking into account the identity of final particles in $p+p\to p+p+V^0$, can be written in terms of the amplitudes ratio:
\begin{equation}
{\cal R}=\displaystyle\frac{\sigma(np\to npV^0)}
{\sigma(pp\to ppV^0)}=\displaystyle\frac{1}{2}+\displaystyle\frac{1}{2}
\displaystyle\frac{|f_{01}|^2}{|f_{10}|^2}.
\label{eq:eqr}
\end{equation}
So a measurement of this ratio (with unpolarized particles) can be considered equivalent to a polarization measurement, as the following relation holds: $
{\cal A}_1=-1+1/(2{\cal R}).$

The polarization transfer from the initial neutron to the final proton 
($\vec n+p\to n+\vec p+V$), can be parameterized as follows:
$
{\cal P}_f=p_1 \vec P_1+p_2 \hat{\vec k} (\hat{\vec k}\cdot\vec P_1),
$
with 
$$p_1(np) =-p_2(np)=\displaystyle\frac{2{\cal R}e f_{01}f_{10}^*}{|f_{01}|^2+|f_{10}|^2}=\cos\delta\displaystyle\frac
{\sqrt{2{\cal R}-1}}{{\cal R}},$$
where $\delta$ is the relative phase of $f_{01}$ and $f_{10}$, which is non zero, in the general case. For the process $p+p\to p+p+V^0$ the relation $p_1(pp) =p_2(pp)=0$ holds, for any vector meson $V^0$.

{\bf What do we learn from the existing data and what COSY can measure?}

DISTO \cite{DISTO} measured the differential cross section for the reaction $p+p\to p+p+\phi$ at $P_{Lab}$=3.67 GeV, at an energy excess $Q=83$ MeV over threshold. The $\phi$ distribution is isotropic, showing that the $\phi$ is emitted in an S-wave state relatively to the two protons. However the $pp$-system can be excited to higher partial wave, as the $pp$ interaction is mainly governed by $\pi$-exchange. The presence of a P-wave in the final $pp$-system, changes the spin structure of the matrix element and induces two transitions, corresponding to singlet initial $pp$ state and $\ell$=0 or 2. In particular, for $\ell$=0, ${\cal M}\propto I\bigotimes\vec \sigma\times\hat{\vec p}\cdot \vec {U^*}$, and $\hat{\vec p}$ is the relative $pp$-momentum. In this case the polarization of the $V$-meson is not transversal, anymore, as it is normal to the momentum of the scattered proton. The angular distribution of the $K$-meson in the $\phi$-reference frame can be parametrized as $$W(\theta_{\phi}^K)\propto \rho_{00}\cos^2(\theta_{\phi}^K)+\rho_{11}\sin^2(\theta_{\phi}^K)
$$
and a fit of the data gives: $ \rho_{00}=0.23\pm 0.04$. This number can be directly related to the ratio between P- and S-wave amplitudes. Within this assumption, one can predict the following values for the spin correlation coefficients: $C_{kk}=0.74$, $C_{nn}=C_{mm}=0.13$, and $A_y=0$ (the simplest case where $A_y$ would not vanish is that the angular momentum of the $\phi$-meson with respect to the $pp$-system is equal to 1. Then $A_y$ would be proportional to the relative phase of these amplitudes).

{\it At COSY it is possible to determine the energy dependence of the P and S waves ratio, in the threshold region, through the measurement of the polarization coefficients and to determine the analyzing power as a function of $\theta_{\phi}^K$ and of the proton emission angle $\theta_p$.}

\subsection{Strange particle production}

The reaction $N+N\to N+Y+K,$ $Y= \Lambda(\Sigma)$ differs from meson production, as the final baryons are not identical. Therefore, even in case of $pp$-reactions, at threshold, two amplitudes are present:
$$ S_i=1, ~\ell_i=1\to {\cal J}^P=0^-\to S_f=0, f_{10}~  \to ~triplet~ to ~singlet$$
$$ S_i=1, ~\ell_i=1\to {\cal J}^P=1^-\to S_f=1, f_{11}~  \to ~triplet~ to ~triplet$$
and the unpolarized cross section can be written as:
$ ({d\sigma}/{d\omega} )_0 =|f_{10}|^2+2|f_{11}|^2$.  The correlation coefficients can be expressed in terms of the two amplitudes and of the unpolarized cross section as: $C_{nn}({d\sigma}/{d\omega} )_0=|f_{10}|^2$ and $C_{kk}({d\sigma}/{d\omega} )_0=-|f_{10}|^2+2|f_{11}|^2$. Due to the symmetry properties of the strong interaction, they are related by a model independent relation: $2C_{nn}+C_{kk}=1$.

{\bf What do we learn from the existing data and what COSY can measure?}

DISTO has measured the polarization transfer coefficient $D_{nn}$  at $P_{Lab}$=3.67 GeV/c and its dependence on the transverse momentum. The values are negative, starting form $\simeq$ -0.6 and
tend towards zero as the momentum increases. From these data it is possible to have a hint of the reaction mechanism, and on the strength and the nature of the $\Lambda N$-interaction. 

$D_{nn}$ is related to the two possible amplitudes by an interference term: $D_{nn}\left ( {d\sigma}{d\omega}\right )_0=2{\cal R}e~f_{10}f_{11}^*$
The reaction can proceed through $K^+$ or $\pi^0$-exchange. In case of one-meson exchange, the two amplitudes are proportional, but with opposite sign:
$$f_{10}^{(K)}=-f_{11}^{(K)}~\mbox{~and~} D_{nn}=-2/3
~\mbox{~for~ K~ exchange},$$
$$f_{10}^{(\pi)}=-f_{11}^{(\pi)}~\mbox{~and~} D_{nn}=2/3
~\mbox{~for~ $\pi$~ exchange};$$
$C_{nn}=C_{kk}=C_{mm}=1/3~\mbox{~for~any~ exchange}.$ For $K+\pi$-exchange, one can express the observables as a function of one parameter, the ratio of the amplitudes for $K^+$ and $\pi^0$-exchange $r=f^{(\pi)}/f^{(K)}$ . In this case one finds: $C_{nn}=(1-2r+r^2)/(3+2r+3r^2)$ and $D_{nn}=2(r^2-2)/(3+2r+3r^2)$, and these observables become sensitive to the relative role of the two mesons  (Fig. 1). The DISTO data ($D_{nn}\simeq -0.4)$ suggest two possible values for $r$: $r^-=-0.64$ and $r^+=0.39$. In this context, 
a measurement of $C_{nn}$ will be discriminative between 
the following values $C_{nn}^-=0.91$ or $C_{nn}^+=0.09$.  

Alternatively, from these data one can extract interesting information on the $\Lambda N$-interaction.
Denoting the ratio between singlet $a_s$ and triplet $a_t$ scattering length as $R_{st}=a_s/a_t$, one can show (for $K$-exchange only) that:
$$ D_{nn}\simeq \displaystyle\frac{-2R_{st}}
{2+R_{st^2}}=-\displaystyle\frac{2}{3}+\Delta,\mbox{~with~}\Delta=\displaystyle\frac{2}{3}\displaystyle\frac{( R_{st}-1)(R_{st}-2)}{2+ R_{st}^2}
$$
DISTO results suggest that $ R_{st}<1$ or $ R_{st}>2$ and the first possibility seems to be in agreement with the last microscopic calculations from J\"ulich \cite{Ha02}.

{\it At COSY it is possible to do the complete experiment at threshold, for $p+p\to p\Lambda K^+$, which requires the measurement of the unpolarized cross section, $C_{nn}$ and $D_{nn}$. Moreover, polarization observables can give an experimental signature of the parity of the the strange particles, through the measurement of the sign of the spin correlation coefficient, $C_{nn}$ or of the polarization transfer coefficient from the initial proton to the produced hyperon \cite{Pa99}.}

\section{Conclusions}

We have illustrated a formalism which allows to get model independent predictions, in the near threshold region, for vector meson and strange particle production. We have shown the importance of polarization observables in further measurement at COSY.

A similar analysis can be done for scalar mesons where the complete experiment in threshold conditions can be realized at COSY, for example for $N+N\to N+N+f_0 (a_0)$.

We thank the organizers of the COSY Summer School and Workshop 2002 for a very lively week, fruitful exchanges and useful discussions.

\begin{figure}
\mbox{\epsfysize=15.cm\leavevmode \epsffile{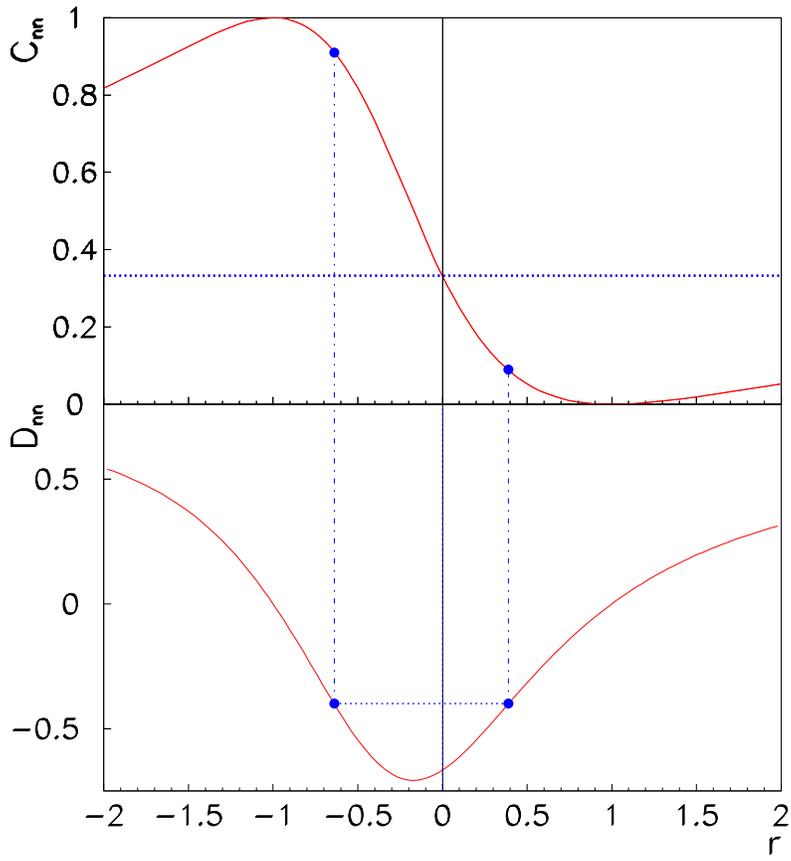}}
\vspace*{.2 truecm}
\caption{ $r=f_{\pi}/f_K$-dependence of $C_{nn}$ (top) and $D_{nn}$ (bottom). The lines indicate the points suggested by DISTO (see text).}
\label{fig:fig1}
\end{figure}

\end{document}